\documentclass[12pt]{article}

\catcode`\@=11

\global\arraycolsep=2pt
\usepackage{amsbsy,amssymb,latexsym,amsfonts,amsmath}
\usepackage{graphicx,color}

\allowdisplaybreaks

\begin{document}

\vspace*{0.7cm}

\begin{center}
{ \Large Dilaton invading from infinitesimal extra dimension}
\vspace*{1.5cm}\\
{Minoru Matsumoto and Yu Nakayama}
\end{center}
\vspace*{1.0cm}
\begin{center}

Department of Physics, Rikkyo University, Toshima, Tokyo 171-8501, Japan

\vspace{3.8cm}
\end{center}

\begin{abstract}
We show that the Wess-Zumino action for the spontaneously broken Weyl (or conformal) symmetry, a.k.a dilaton effective action, in even $D$ dimensions can be obtained from the Kaluza-Klein dimensional reduction of the Lovelock action in $D+\epsilon$ dimensions by taking the $\epsilon  \to 0$ limit, where the dilaton is identified with the metric in the extra dimension. 
The construction gives an explicit form of the dilaton effective action in any even dimensions.
\end{abstract}

\thispagestyle{empty} 

\setcounter{page}{0}

\newpage

\section{Introduction}

The Wess-Zumino action for the spontaneously broken Weyl (or conformal) symmetry, a.k.a dilaton effective action, plays a significant role in our understanding of the renormalization group (see \cite{Komargodski:2011xv}\cite{Nakayama:2013is}\cite{Shore:2016xor} for reviews). It is constructed such that its Weyl variation reproduces the Weyl anomaly (or conformal anomaly), but the explicit form of the dilaton effective action is highly non-trivial. It has some mathematical applications in conformal geometries in relation to the higher-dimensional conformal operators \cite{confL} and Q-curvatures \cite{Q}.

In two dimensions, the dilaton effective action is the Liouville theory and it has been long studied in the context of two-dimensional gravity and worldsheet (non-critical) string theory \cite{Polyakov:1981rd}. In four dimensions, the explicit form first appeard in \cite{Fradkin:1983tg}, and the positivity constraints on the dilaton effective action lead to the proof of the $a$-theorem \cite{Komargodski:2011vj}\cite{Luty:2012ww}, which is a milestone of four-dimensional quantum field theories. The applications of the dilaton effective action in six dimensions have been pursued in the more recent literature \cite{Elvang:2012st}\cite{Coriano:2013nja}\cite{Cordova:2015fha}\cite{Heckman:2021nwg}.

In this paper we show that the dilaton effective action can be obtained from the Kaluza-Klein dimensional reduction of the Lovelock action in $D+\epsilon$ dimensions by taking the $\epsilon \to 0$ limit, where the dilaton is identified with the metric in the extra dimension. One advantage of this approach is that we can give an explicit form of the dilaton effective action in any even dimensions.\footnote{A related observation in four-dimensions was made in the recent paper \cite{Coriano:2022knl}. The dimensional reduction approach to obtain modified theories of gravity (a.k.a ``four-dimensional Gauss-Bonnet gravity") was introduced in \cite{Lu:2020iav}\cite{Kobayashi:2020wqy}\cite{Hennigar:2020lsl}\cite{Easson:2020mpq}. See also \cite{Baume:2013ika}\cite{Herzog:2015ioa} for the similar approach.} They may be useful in understanding the structure of the renormalization group in higher dimensions. 

The rest of the paper is organized as follows. In section 2, we derive the Wess-Zumino action for the spontaneously broken Weyl symmetry from the Kaluza-Klein dimensional reduction of the Lovelock gravity in $D+\epsilon$ dimensions. In section 3, we show some explicit expressions of the dilaton effective action. In section 4, we conclude with some discussions.

\section{Dilaton effective action from Kaluza-Klein reduction}

Let us consider $D$-dimensional quantum field theories coupled with a background metric $g_{\mu\nu}$. In even dimensions, the Weyl anomaly \cite{Deser:1976yx}\cite{Duff:1977ay}\cite{Deser:1993yx} takes the form of
\begin{align}
\langle T^\mu_{\ \mu} \rangle = a_D \mathrm{Euler}_D + \sum_i c_i \mathrm{Weyl}_i \ ,
\end{align}
up to trivial terms that can be removed by local counter-terms. Here, $\mathrm{Euler}_D$ is the Euler density in (even) $D$ dimensions, and $\mathrm{Weyl}_i$ represent various curvature invariants of order $D/2$ constructed out of the Weyl tensor $C_{\mu\nu\rho\sigma}$.\footnote{In two dimensions, there is none; in four dimensions, there is one CP even invariant $C_{\mu\nu\rho\sigma}C^{\mu\nu\rho\sigma}$ and one CP odd invariant $\epsilon^{\mu\nu \alpha \beta}C_{\mu\nu\rho\sigma}C_{\alpha \beta}^{\ \ \ \ \rho\sigma}$. In higher dimensions, there exist more invariants. For the discussions on CP odd invariants, see e.g. \cite{Nakayama:2012gu}\cite{Nakagawa:2020gqc} and reference therein.}

Our goal is to construct the Wess-Zumino effective action $S[g_{\mu\nu},\tau]$ such that the compensated Weyl transformation $g_{\mu\nu} \to e^{2\sigma}g_{\mu\nu}$ and $\tau \to \tau+\sigma$ reproduces the Weyl anomaly
\begin{align}
\delta S = \int d^D x \sqrt{g} \delta \sigma (a_D \mathrm{Euler}_D + \sum_i c_i \mathrm{Weyl}_i) \ .
\end{align}
The scalar $\tau$ is known as a dilaton: when the Wess-Zumino action is induced by the spontaneous breaking of Weyl (or conformal) symmetry, it plays a role of the Nambu-Goldstone boson.

It is trivial to construct the dilaton effective action for the Weyl invariant terms in the Weyl anomaly (so-called type A anomaly):
\begin{align}
S_{\mathrm{type A}} = \int d^D x \sqrt{g} \tau \sum_i c_i \mathrm{Weyl}_i ,
\end{align}
but it has been a non-trivial task to determine the dilaton effective action for the Euler term (type B anomaly) in the Weyl anomaly. The brute-force Noether method and the Wess-Zumino trick (i.e. extending the action in $D+1$ dimensions with boundary) was studied in the literature (e.g. \cite{Elvang:2012st}).

Our claim is that the dilaton effective action (for type B anomaly) can be efficiently obtained by studying the Kaluza-Klein dimensional reduction of the of the Lovelock action in $D+\epsilon$ dimensions by taking the $\epsilon  \to 0$ limit.
\begin{align}
S_{\mathrm{type B}} = \int_{X^D} d^Dx \sqrt{g_D} L_D = \lim_{\epsilon \to 0} \frac{1}{2^{D/2}\epsilon} \int_{X^D \times T^\epsilon} d^{D+\epsilon} x \sqrt{g_{D+\epsilon}} L_{(D/2)}^{D+\epsilon} \ , \label{action}
\end{align}
where $L_{(D/2)}^{D+\epsilon}$ is the Lovelock Lagrangian of order $D/2$ in $D+\epsilon$ dimensions \cite{Lovelock:1971yv}:
\begin{align}
L_{(D/2)}^{D+\epsilon} = \delta^{A_1 \cdots A_D}_{[B_1\cdots B_D]}R_{A_1 A_2}^{\ \ \ \ \ B_1 B_2} \cdots R_{A_{D-1}A_{D}}^{\ \ \ \ \  \ \ B_{D-1} B_D} \ .   
\end{align}
Note that the Lovelock Lagrangian does not depend on $\epsilon$ explicitly (but it is implicitly affected through the index summation). We also note that $L_{(D/2)}^{D} = 2^{D/2} \mathrm{Euler}_D$.

Under the Kaluza-Klein dimensional reduction, we assume the following form of the metric:
\begin{align}
ds_{D+\epsilon}^2 = g_{\mu\nu} dx^\mu dx^\nu + e^{2\tau} ds^2_{T^\epsilon} \ , \label{KKmetric}
\end{align}
where $ds^2_{T^\epsilon}$ is the (locally) flat metric in $\epsilon$ dimensions. In practice we study the Kaluza-Klein compactification of the $N$ dimensional torus and analytically continue the dimension $N \to \epsilon$ and take the limit $\epsilon \to 0$. In this limit, we keep the order $\epsilon$ terms to cancel the $1/\epsilon$ factor in  \eqref{action}, which will result in the non-trivial $\tau$ dependence in $D$ dimensions.

Before presenting the explicit formulae, we first provide a proof that this construction gives the desired (compensated) Weyl variation that reproduces the Weyl anomaly. A key observation is that the compensated Weyl variation (i.e. $g_{\mu\nu} \to e^{2\sigma}g_{\mu\nu}$ and $\tau \to \tau+\sigma$ in $D$ dimensions) induces the Weyl variation of the $D+\epsilon$ dimensional metric $ds_{D+\epsilon}^2 \to e^{2\sigma} ds_{D+\epsilon}^2$. This observation will justify our particular choice of the dilaton dependence in the Kaluza-Klein ansatz \eqref{KKmetric}.

Investigating the behavior under the constant Weyl variation of the $D+\epsilon$ dimensional Lovelock action and then reducing the variation to $D$ dimensions, we see that the variation under the general Weyl variation should take the form
\begin{align}
\delta S = \frac{1}{\epsilon} \int_{X^D} d^{D} x \sqrt{g_D}(\epsilon \delta\sigma(\mathrm{Euler}_D + O(\epsilon)) +J^\mu \partial_\mu \delta\sigma ) \  . 
\end{align}
To obtain the desired Weyl variation,  we need to assure that the local vector operator $J_\mu$, which could depend on $\tau$, vanishes as fast as $O(\epsilon^2)$ (rather than $O(\epsilon)$). This follows from the crucial observation that the metric variation of the Lovelock action gives only terms with no derivative acting on curvature tensors. By assuring $J_\mu =  O(\epsilon^2)$, the resultant variation in the $\epsilon \to 0$ limit is precisely type B Weyl anomaly i.e. the Euler density in $D$ dimensions.

 
We should note that if we started with arbitrary diffeomorphism invariant action in $D+\epsilon$ dimensions rather than the Lovelock action, the Kaluza-Klein dimensional reduction would not give the dilaton effective action because the Weyl variation would contain derivatives of $\tau$ generally. It is the crucial property of the Lovelock action that leads to the desired results.\footnote{To avoid confusion, we stress that this property is sufficient but not necessary. The Wess-Zumino action for (CP even) type A Weyl anomaly can be obtained from the Kaluza-Klein dimensional reduction of the curvature invariants constructed out of Weyl tensors in $D+\epsilon$ dimensions although {\it general} metric variation gives higher derivative equations of motion.}
 
 With this proof, we now give the explicit formulae. The Kaluza-Klein dimensional reduction of the Lovelock acion in $D+N$ dimensions, where $N$ is an integer, can be found in \cite{VanAcoleyen:2011mj}:\footnote{Our dilaton is related to their Galileon field by $\pi = 2\tau$.}
 \begin{align}
S_N = \int_{X^D} d^D x \sqrt{g_D} e^{N\tau}\left( L_{(D/2)}^D + \sum_{n=1}^{D-1} \hat{C}^n \mathcal{L}^n \right) \ ,
 \end{align}
 where 
 \begin{align}
\hat{C}^n &= \frac{D-n-N}{(D-n)(D-n+1)} \frac{(-2)^{2n-D/2}(D/2)! N!} {2 (N+n-D)!}  \cr
\mathcal{L}^n&= \sum_{p= \mathrm{max}(0,n-D/2)}^{\frac{n-1}{2}} \mathcal{C}^n_p A_{(2n)} (4\partial_{\mu_1} \tau \partial_{\mu_2} \tau) \mathcal{S}(q) (4\partial_\rho \tau \partial^\rho \tau)^{D/2 -n} \mathcal{R}(p) \cr 
\mathcal{C}^n_{p}&=\frac{(-2)^{-3p}}{p!(D/2-n+p)!(n-2p-1)!} \cr
\mathcal{S}(q) &= \prod_{i=a}^{q+a-1}D_{\mu_{2i-1}}\partial_{\mu_{2i}}(2\tau) \cr
\mathcal{R}(p) &= (4\partial_\rho \tau \partial^\rho \tau)^p \prod_{k=b}^{p+b-1} R_{\mu_{4k-3}\mu_{4k-1} \mu_{4k-2}\mu_{4k}} \cr
A_{(2n)} & = \delta^{\mu_1 \mu_3 \cdots \mu_{2n-1}}_{[\nu_2 \nu_4 \cdots \nu_{2n}]} g^{\nu_2 \mu_2} \cdots g^{\nu_{2n} \mu_{2n}}  \ , \label{notation}
 \end{align}
 where $q=n-1-2p$. Schematically, $\mathcal{L}^n \sim \sum_p (\Box \tau)^q (\partial \tau \partial \tau)^{D/2-n+p+1} R^p$, so the power of $\tau$ in each term $\mathcal{L}^n$ is fixed by the summation index $n$ as $D-n+1$.
 
 By analytically continuing $N \to \epsilon$ (i.e. in the denominator we have $(N+n-D)! = \Gamma(N + n- D +1) \to \Gamma(\epsilon + n - D + 1)  \sim \frac{1}{\epsilon} \frac{(-1)^{D-n-1}}{(D-n-1)!}$ ) and taking the $\epsilon \to 0$ limit, we obtain
\begin{align}
S_{\mathrm{type B}} &= \int_{X^D} d^D x \sqrt{g_D}\left( \tau \mathrm{Euler}_{D}+ \sum_{n=1}^{D-1} \hat{C}^n_r \mathcal{L}^n \right)   \cr
\hat{C}^n_r &= \frac{(-1)^{D-n-1} (-2)^{2n-D} (D-n-1)! (D/2)!}{2(D-n+1) } \ . 
\end{align}

In particular, the dilaton effective action evaluated in the flat space-time takes the form
\begin{align}
S_{\mathrm{flat}} = \int d^D x  \sum_{n=1}^{D/2} \hat{C}^n_r \mathcal{L}^n(p=0) \ ,
\end{align}
where the summation is only over $p=0$ term (i.e. $q=n-1$) in $\mathcal{L}^n$. The summation over $n$ is only up to $D/2$ and lower powers of $\tau$ vanish.

When we compare our formulae with the results in the literature, we may have to take into account two things. The first is that the dilaton effective action is determined only up to (compensated) Weyl invariant terms (i.e. curvature invariants constructed out of the Weyl compensated metric $e^{-2\tau}g_{\mu\nu}$). The resulting expression may be different by these terms, which does not affect the (compensated) Weyl variation.\footnote{Our scheme is such that we have as many $\tau$ as possible. This is in contrast with the $Q$-curvature approach. See e.g. \cite{Mazur:2001aa}\cite{Elvang:2012yc}.} The second is that the Lagrangian may look different by total derivatives with the help of the Bianchi identity and the dimension-specific curvature identities. 

\section{Explicit form of the dilaton effective action}
For reference, we collect the explicit form of the dilaton effective action in $D=2,4,6$. In $D=2$, we have 
\begin{align}
\mathcal{L}^1 &= 4(\partial^\rho \tau \partial_\rho \tau) 
\end{align}
so that we obtain
\begin{align}
S_2 = \int d^2x \sqrt{g} (\tau \mathrm{Euler}_2 {-} \partial_\mu \tau \partial^\mu \tau) \ , 
\end{align}
where $\mathrm{Euler}_2 = R$.

In $D=4$, we have 
\begin{align}
\mathcal{L}^1 &= 16(\partial^\rho \tau \partial_\rho \tau)^2 \cr
\mathcal{L}^2 &= 8((\partial^\rho \tau \partial_\rho \tau) \Box \tau - \partial_\mu \tau \partial_\nu \tau D^\mu \partial^\nu \tau) \cr \mathcal{L}^3 &= 2 \partial_\mu \tau \partial_\nu \tau G^{\mu\nu}
\end{align}
so that up on integration by part (in $\mathcal{L}^2$), we obtain
\begin{align}
S_4 =  \int d^4x \sqrt{g} (\tau \mathrm{Euler}_4 + 4(R_{\mu\nu}-\frac{1}{2}R g_{\mu\nu}) \partial^\mu \tau \partial^\nu \tau  -4(\partial_\rho \tau \partial^\rho \tau) \Box \tau + 2(\partial_\rho \tau \partial^\rho \tau)^2) \ , 
\end{align}
where $\mathrm{Euler}_4 = R_{\mu\nu\rho\sigma}R^{\mu\nu\rho\sigma} - 4 R_{\mu\nu}R^{\mu\nu} + R^2$.

In $D=6$, we have 
\begin{align}
\mathcal{L}^1 &= 64 \mathcal{C}_0^1(\partial^\rho \tau \partial_\rho \tau)^3 \cr
\mathcal{L}^2 &= 32 \mathcal{C}_0^2 ((\partial^\rho \tau \partial_\rho \tau) \Box \tau - \partial_\mu \tau \partial_\nu \tau D^\mu \partial^\nu \tau) (\partial^\rho \tau \partial_\rho \tau) \cr 
\mathcal{L}^3 &= 16\mathcal{C}_0^3 \left( (\partial_\rho \tau \partial^\rho \tau)( (\Box \tau)^2 -D^\mu \partial^\nu \tau D_\mu \partial_\nu \tau) -2\partial_\mu \tau \partial_\nu\tau(D^\mu\partial^\nu\tau \Box\tau - D^\mu \partial_\rho \tau D^\nu \partial^\rho \tau) \right) \cr 
&+ 16\mathcal{C}_1^3 (2 (\partial^\rho \tau \partial_\rho \tau) R -4 \partial_\mu\tau \partial_\nu \tau R^{\mu\nu})(\partial^\sigma \tau \partial_\sigma \tau) \cr
\mathcal{L}^4 &= 8 \mathcal{C}_1^4 (2 (\partial^\rho \tau \partial_\rho \tau) \Box \tau -2\partial_\mu \tau \partial_\nu \tau D^\mu \partial^\nu \tau)R \cr
 &-4((\partial^\rho \tau \partial_\rho \tau) D^\mu \partial^\nu \tau -2 \partial_\rho \tau \partial^\mu\tau  D^{\rho}\partial^\nu \tau + \partial^\mu \tau \partial^\nu \tau \Box \tau )R_{\mu\nu} +4(\partial_\mu \tau \partial_\rho\tau D_\nu \partial_\sigma \tau ) R^{\mu\nu\rho\sigma}) \cr
\mathcal{L}^5 &= 4 \mathcal{C}_2^5 (4 \mathrm{Euler}_4 (\partial^\rho \tau \partial_\rho \tau) \cr 
 & -16(R^{\rho\sigma \lambda \nu}R_{\rho\sigma \lambda \mu} + 2 R_{\lambda \mu \ \sigma}^{\ \ \ \nu} R^{\sigma \lambda}-2R_{\sigma \mu} R^{\sigma \nu}+ R R_{\mu}^{\ \nu} ) \partial^\mu \tau \partial_\nu \tau )  
\end{align}
so that up on integrating by part (in $\mathcal{L}^2$ and  $\mathcal{L}^3$), we obtain
\begin{align}
S_6 = \int d^6x \sqrt{g} &(\tau \mathrm{Euler}_6 -3 \mathrm{Euler}_4 \partial_\rho \tau \partial^\rho \tau  + 12 R_{\mu\rho\sigma \lambda}R^{\nu \rho\sigma\lambda} \partial^\mu\tau \partial_\nu \tau \cr 
& -24 R^{\mu\rho\nu\sigma} R_{\rho\sigma} \partial_\mu\tau \partial_\nu \tau -24 R^{\mu\rho} R_{\nu\rho} \partial_\mu\tau \partial^\nu \tau + 12 R R^{\mu\nu} \partial_\mu \tau \partial_\nu \tau 
\cr
 &-8 ((\partial^\rho \tau \partial_\rho \tau) \Box \tau - \partial_\mu \tau \partial_\nu \tau D^\mu \partial^\nu \tau)R -16(\partial_\mu \tau \partial_\rho\tau D_\nu \partial_\sigma \tau ) R^{\mu\nu\rho\sigma}  \cr
 &+16((\partial^\rho \tau \partial_\rho \tau) D^\mu \partial^\nu \tau -2 \partial_\rho \tau \partial^\mu\tau  D^{\rho}\partial^\nu \tau + \partial^\mu \tau \partial^\nu \tau \Box \tau )R_{\mu\nu} 
\cr 
&+6R(\partial_\rho\tau \partial^\rho\tau)^2  + 24(\partial^\rho \tau \partial_\rho \tau)D^{\mu}\partial^\nu \tau D_{\mu}\partial_\nu \tau -24(\partial^\rho \tau \partial_\rho \tau)(\Box \tau)^2  \cr
 &+36 (\partial^\rho \tau \partial_\rho \tau)^2(\Box \tau) -24 (\partial^\rho \tau \partial_\rho \tau)^3) \ , 
\end{align}
where
\begin{align}
\mathrm{Euler}_6 &= R^3-12R R_{\mu\nu}R^{\mu\nu} + 3R R_{\mu\nu\rho\sigma}R^{\mu\nu\rho\sigma} \cr 
&-24 R^{\mu \lambda}_{\ \ \rho \sigma}R^{\rho \sigma \  \nu}_{\ \ \mu} R_{\nu\lambda} + 24 R_{\mu \nu}R^{\mu\rho\nu\sigma} R_{\sigma\rho} + 16 R_{\mu}^{\ \nu} R_{\nu}^{\ \rho} R_{\rho}^{\ \mu}   \cr 
&+2 R^{\mu\nu}_{\ \ \rho\sigma} R^{\rho\sigma}_{\ \ \lambda \omega} R^{\lambda \omega}_{\ \ \mu\nu} - 8 R^{\mu \nu}_{\ \ \rho\sigma} R_{\mu \lambda}^{\ \ \rho\omega} R_{\nu \omega}^{\ \ \sigma \lambda}  \ .
\end{align}
They all agree with the ones in the literature \cite{Komargodski:2011vj}\cite{Elvang:2012st}.\footnote{The  powers of curvature tensors in $D=6$ dimensions admit various identities to rewrite the Euler density. See e.g. Appendix of \cite{Decanini:2007gj}.}

In order to compare our expression with that in \cite{Baume:2013ika} more directly, the following identity found in Appendix of \cite{VanAcoleyen:2011mj} may be useful.
\begin{align}
\frac{D-n}{2}\mathcal{K}^n -2\mathcal{L}^{n+1} = D^\mu J^n_\mu 
\end{align}
with a local vector field $J^n_\mu$. Here, 
\begin{align}
\mathcal{K}^n &= \sum_p \mathcal{D}_p^n A_{(2n)}\mathcal{S}(q+1)(4\partial_\rho\tau \partial^\rho \tau)^{D/2-n} \mathcal{R}(p) \cr
\mathcal{D}_p^n &= \frac{2}{n-p} \mathcal{C}_p^n \ . 
\end{align}
See \eqref{notation} for the notation.
The identity allows us to replace $\mathcal{L}^{n+1}$ with $\mathcal{K}^n$  inside the dilaton effective action.

\section{Discussions}
In this note, we have shown that the dilaton effective action can be obtained  from the Kaluza-Klein dimensional reduction of the Lovelock action in $D+\epsilon$ dimensions by taking the $\epsilon \to 0$ limit. The same is true for (CP even) type A Weyl anomaly if we perform the Kaluza-Klein reduction of the action constructed out of the Weyl tensors in $D+\epsilon$ dimensions (by subtracting the $O(\epsilon^{-1})$ Weyl invariant terms in $D$ dimensions). In this way, we can obtain all the relevant terms in the dilaton effective action from the Kaluza-Klein dimensional reduction.\footnote{For the CP odd type A anomaly, the dimensional continuation of the Levi-Civita tensor may be non-trivial.} In our approach, the dilaton is identified with a component of the metric in the extra dimension.

At the technical level, we have given the proof of how the Kaluza-Klein dimensional reduction works. Nonetheless, it remains a little mysterious   why we have to start with these particular actions in $D+\epsilon$ dimensions. The Lovelock action has a crucial property that the metric variation does not contain derivatives on curvature, but the relevance of this property, in particular, if we regard our $D+\epsilon$ dimensional setup as a dimensional regularization of the quantum field theory, is not immediately clear. We also would like to understand the origin of the (compensated) Weyl invariant counter-terms in our dimensional reduction approach.

Our results may be useful in understanding constraints on the renormalization group flow in higher-dimensional quantum field theories. In order to provide a universal constraint, we have to assure that the above-mentioned counter-terms would not affect the physical predictions of the dilaton effective action, which should be confirmed in future studies.

Finally, our results show that the origin of the so-called ``four-dimensnoal Gauss-Bonnet gravity"  (see e.g. \cite{Glavan:2019inb}\cite{Lu:2020iav}\cite{Kobayashi:2020wqy}\cite{Hennigar:2020lsl}\cite{Coriano:2022knl} and reference therein) may be the dilaton associated with the spontaneously broken Weyl (conformal) symmetry. It is quite interesting if the dilaton exists in nature and our cosmology is governed by the spontaneously broken Weyl (conformal) symmetry.

\section*{Acknowledgements}
We would like to thank T.~Kobayashi for comments and the introduction to his work.
This work by YN is in part supported by JSPS KAKENHI Grant Number 21K03581.


\begin{thebibliography}{99}


\bibitem{Komargodski:2011xv}
Z.~Komargodski,
JHEP \textbf{07}, 069 (2012)
doi:10.1007/JHEP07(2012)069
[arXiv:1112.4538 [hep-th]].


\bibitem{Nakayama:2013is}
Y.~Nakayama,
Phys. Rept. \textbf{569}, 1-93 (2015)
doi:10.1016/j.physrep.2014.12.003
[arXiv:1302.0884 [hep-th]].


\bibitem{Shore:2016xor}
G.~M.~Shore,
doi:10.1007/978-3-319-54000-9
[arXiv:1601.06662 [hep-th]].



\bibitem{confL}
C.R. Graham, R. Jenne, L.J. Mason and G.A.J. Sparling, J. London Math. Soc. (2) 46, 557 (1992).


\bibitem{Q}
T.~Branson, Math.Scand. 57 (1985) 293


\bibitem{Polyakov:1981rd}
A.~M.~Polyakov,
Phys. Lett. B \textbf{103}, 207-210 (1981)
doi:10.1016/0370-2693(81)90743-7

\bibitem{Fradkin:1983tg}
E.~S.~Fradkin and A.~A.~Tseytlin,
Phys. Lett. B \textbf{134}, 187 (1984)
doi:10.1016/0370-2693(84)90668-3


\bibitem{Komargodski:2011vj}
Z.~Komargodski and A.~Schwimmer,
JHEP \textbf{12}, 099 (2011)
doi:10.1007/JHEP12(2011)099
[arXiv:1107.3987 [hep-th]].

\bibitem{Luty:2012ww}
M.~A.~Luty, J.~Polchinski and R.~Rattazzi,
JHEP \textbf{01}, 152 (2013)
doi:10.1007/JHEP01(2013)152
[arXiv:1204.5221 [hep-th]].


\bibitem{Elvang:2012st}
H.~Elvang, D.~Z.~Freedman, L.~Y.~Hung, M.~Kiermaier, R.~C.~Myers and S.~Theisen,
JHEP \textbf{10}, 011 (2012)
doi:10.1007/JHEP10(2012)011
[arXiv:1205.3994 [hep-th]].

\bibitem{Coriano:2013nja}
C.~Coriano, L.~Delle Rose, C.~Marzo and M.~Serino,
Class. Quant. Grav. \textbf{31} (2014), 105009
doi:10.1088/0264-9381/31/10/105009
[arXiv:1311.1804 [hep-th]].


\bibitem{Cordova:2015fha}
C.~Cordova, T.~T.~Dumitrescu and K.~Intriligator,
JHEP \textbf{10}, 080 (2016)
doi:10.1007/JHEP10(2016)080
[arXiv:1506.03807 [hep-th]].


\bibitem{Heckman:2021nwg}
J.~J.~Heckman, S.~Kundu and H.~Y.~Zhang,
Phys. Rev. D \textbf{104}, no.8, 085017 (2021)
doi:10.1103/PhysRevD.104.085017
[arXiv:2103.13395 [hep-th]].

\bibitem{Coriano:2022knl}
C.~Corian\`o and M.~M.~Maglio,
[arXiv:2201.07515 [hep-th]].







\bibitem{Lu:2020iav}
H.~Lu and Y.~Pang,
Phys. Lett. B \textbf{809}, 135717 (2020)
doi:10.1016/j.physletb.2020.135717
[arXiv:2003.11552 [gr-qc]].

\bibitem{Kobayashi:2020wqy}
T.~Kobayashi,
JCAP \textbf{07}, 013 (2020)
doi:10.1088/1475-7516/2020/07/013
[arXiv:2003.12771 [gr-qc]].


\bibitem{Hennigar:2020lsl}
R.~A.~Hennigar, D.~Kubiz\v{n}\'ak, R.~B.~Mann and C.~Pollack,
JHEP \textbf{07}, 027 (2020)
doi:10.1007/JHEP07(2020)027
[arXiv:2004.09472 [gr-qc]].



\bibitem{Easson:2020mpq}
D.~A.~Easson, T.~Manton and A.~Svesko,
JCAP \textbf{10}, 026 (2020)
doi:10.1088/1475-7516/2020/10/026
[arXiv:2005.12292 [hep-th]].


\bibitem{Baume:2013ika}
F.~Baume and B.~Keren-Zur,
JHEP \textbf{11}, 102 (2013)
doi:10.1007/JHEP11(2013)102
[arXiv:1307.0484 [hep-th]].

\bibitem{Herzog:2015ioa}
C.~P.~Herzog, K.~W.~Huang and K.~Jensen,
JHEP \textbf{01}, 162 (2016)
doi:10.1007/JHEP01(2016)162
[arXiv:1510.00021 [hep-th]].


\bibitem{Deser:1976yx}
S.~Deser, M.~J.~Duff and C.~J.~Isham,
Nucl. Phys. B \textbf{111}, 45-55 (1976)
doi:10.1016/0550-3213(76)90480-6

\bibitem{Duff:1977ay}
M.~J.~Duff,
Nucl. Phys. B \textbf{125}, 334-348 (1977)
doi:10.1016/0550-3213(77)90410-2

\bibitem{Deser:1993yx}
S.~Deser and A.~Schwimmer,
Phys. Lett. B \textbf{309}, 279-284 (1993)
doi:10.1016/0370-2693(93)90934-A
[arXiv:hep-th/9302047 [hep-th]].


\bibitem{Nakayama:2012gu}
Y.~Nakayama,
Nucl. Phys. B \textbf{859}, 288-298 (2012)
doi:10.1016/j.nuclphysb.2012.02.006
[arXiv:1201.3428 [hep-th]].

\bibitem{Nakagawa:2020gqc}
K.~Nakagawa and Y.~Nakayama,
Phys. Rev. D \textbf{101}, no.10, 105013 (2020)
doi:10.1103/PhysRevD.101.105013
[arXiv:2002.01128 [hep-th]].




\bibitem{Lovelock:1971yv}
D.~Lovelock,
J. Math. Phys. \textbf{12}, 498-501 (1971)
doi:10.1063/1.1665613

\bibitem{VanAcoleyen:2011mj}
K.~Van Acoleyen and J.~Van Doorsselaere,
Phys. Rev. D \textbf{83}, 084025 (2011)
doi:10.1103/PhysRevD.83.084025
[arXiv:1102.0487 [gr-qc]].


\bibitem{Mazur:2001aa}
P.~O.~Mazur and E.~Mottola,
Phys. Rev. D \textbf{64}, 104022 (2001)
doi:10.1103/PhysRevD.64.104022
[arXiv:hep-th/0106151 [hep-th]].

\bibitem{Elvang:2012yc}
H.~Elvang and T.~M.~Olson,
JHEP \textbf{03}, 034 (2013)
doi:10.1007/JHEP03(2013)034
[arXiv:1209.3424 [hep-th]].


\bibitem{Decanini:2007gj}
Y.~Decanini and A.~Folacci,
Class. Quant. Grav. \textbf{24}, 4777-4799 (2007)
doi:10.1088/0264-9381/24/18/014
[arXiv:0706.0691 [gr-qc]].


\bibitem{Glavan:2019inb}
D.~Glavan and C.~Lin,
Phys. Rev. Lett. \textbf{124}, no.8, 081301 (2020)
doi:10.1103/PhysRevLett.124.081301
[arXiv:1905.03601 [gr-qc]].

\end{thebibliography}
\end{document}